\documentclass[12pt,a4paper]{article}
\usepackage{a4wide}

\usepackage{amsfonts, amscd, mathrsfs, amsopn, bbm, bbold, multicol, relsize}
\usepackage{mathrsfs}
\usepackage[psamsfonts]{amssymb}
\usepackage{mathbbol}
\usepackage[mathscr]{eucal}
\usepackage[cmex10]{amsmath}
\usepackage[amsmath,thmmarks]{ntheorem}
\usepackage{mathtools}
\usepackage{cite}
\usepackage{tikz}
\usepackage{graphicx}

\theoremstyle{plain}
\theoremheaderfont{\itshape\bfseries}
\theorembodyfont{\normalfont\itshape}
\theoremseparator{:}
\theoremsymbol{}

\theoremstyle{nonumberplain}

\theoremstyle{nonumberplain}
\theoremheaderfont{\itshape}
\theorembodyfont{\normalfont}
\theoremseparator{:}
\theoremsymbol{}

\theoremstyle{plain}
\theoremheaderfont{\itshape\bfseries}
\theorembodyfont{\normalfont}
\theoremseparator{:}

\theoremstyle{nonumberplain}
\theoremsymbol{\rule{1ex}{1ex}}

\newlength\fheight
\newlength\fwidth
\def\S{ \mathcal{S} }               

\newcommand\restr[2]{{
  \left.\kern-\nulldelimiterspace 
  #1 
  \right|_{#2} 
  }}

\usepackage[section]{placeins}
\usepackage{url}
\usepackage{epstopdf}
\usepackage{enumerate}
\usepackage{algorithmicx}
\usepackage{algorithm}
\usepackage{amsmath}
\usepackage[noend]{algpseudocode}
\usepackage{float}
\usepackage{color}
\usepackage{makeidx}
\usepackage{bbm}
\usepackage{graphicx}
\usepackage{lipsum}
\usepackage{subcaption}
\usepackage[section]{placeins}
\usepackage{setspace}
\usepackage{lipsum}
\usepackage{adjustbox}

\def\BibTeX{{\rm B\kern-.05em{\sc i\kern-.025em b}\kern-.08em
    T\kern-.1667em\lower.7ex\hbox{E}\kern-.125emX}}

\newcounter{RonCounter}

\newcounter{CagkanCounter}

\begin{document}
\title{Real-time Localization Using Radio Maps}


	\author{\c{C}a\u{g}kan Yapar$^{\ddagger }$\footnote{Equal contribution.} \quad Ron Levie$^{\dagger*}$ \quad  Gitta Kutyniok$^{\dagger\S}$ \quad Giuseppe Caire$^{\ddagger}$ \\
		$^{\ddagger}$ Institute of Telecommunication Systems, 
		TU Berlin,\\
		$^{\dagger}$ Institute of Mathematics,  
		TU Berlin,\\		
		$^{\S}$Department of Physics and Technology, University of Troms{\o}}
	\date{}		

\maketitle

\begin{abstract}
This paper deals with the problem of localization in a cellular network in a dense urban scenario.  Global Navigation Satellite System typically performs poorly in urban environments when there is no line-of-sight between the devices and the satellites, and thus alternative localization methods are often required. We present a simple yet effective method for localization based on pathloss. In our approach, the user to be localized reports the received signal strength from a set of base stations with known locations. For each base station we have a good approximation of the pathloss at each location in the map, provided by RadioUNet, an efficient deep learning-based simulator of pathloss functions in urban environment, akin to ray-tracing.  
Using the approximations of the pathloss functions of all base stations and the reported signal strengths,
we are able to extract a very accurate approximation of the location of the user. 

\end{abstract}

\section{Introduction} \label{sec:Intro}
The location information of a User Equipment (UE) is essential for many current and envisioned applications that range from emergency 911 services \cite{spect}, autonomous driving \cite{autonomousDriving}, intelligent transportation systems \cite{beyondGNSS}, or social networks, asset tracking, advertising, to name some \cite{chal}.

In urban environments, Global Navigation Satellite System (GNSS) fails to provide a reliable localization estimate due to the lack of line-of-sight link between the UE and the GNSS satellites \cite{GPSVehicle}. In addition, usage of GNSS increases power consumption at UE.
It is thus important to use additional fingerprint information to successfully localize the UE. 
In cellular-network based localization, the position of UE can be estimated by using different metrics that UE may report or the network can infer. The prominent localization methods in the literature are based on Time of Arrival (ToA) \cite{TOA1,TOA2}, Time Difference of Arrival (TDoA) \cite{TDOA1}, Angle of Arrival (AoA) \cite{AOA1} and Received Signal Strength (RSS) \cite{RSS1} measurements. RSS-based methods distinguish themselves from other methods by their availability at any device, whereas time-based (ToA and TDoA) and angle-based (AoA) methods require high precision clocks and antenna arrays, respectively \cite{DBCorr,RSStoRange}.

However, most previous RSS works oversimplify the model of the wireless environment, making it inappropriate for urban settings. In the so-called range-based techniques, the measurements are firstly used to estimate the distance between the UE and infrastructure elements such as base stations (BS) \cite{RSStoRange}. Using the signal attenuation (and the resulting measured signal strength) straightforwardly to estimate the distance between the UE and the BS is not appropriate, since in practice the signal undergoes complicated propagation phenomena such as reflections, diffractions, and wave guide effects. As a result, the estimated ranges are inherently flawed. Several publications \cite{NLOSMit,TOA1,TOA2} proposed methods to mitigate the effects of obstructions in range estimation mismatch.  Nevertheless, these methods do not directly use a complete model of the propagation phenomena, and only partially alleviate the aforementioned problem.

In comparison to the methods mentioned above, in our approach we take into account the physical phenomenon  of signal strength/pathloss with no major simplifications. Note that the signal strength profoundly depends on the geometry of the urban environment. In principle, the pathloss function can be computed from the geometry of the urban environment using a physical simulation like ray-tracing \cite{RayTracing,WinPropFEKO}. However, such simulations are too slow and computationally demanding in real time. Moreover, in physical simulations the accurate domain must be known, including the accurate shapes and locations of all buildings, and all other obstacles (like cars). Without an accurate knowledge of the domain, physical simulations do not achieve high accuracy.
Some previous publications that used ray-tracing simulations for localization include \cite{FPWinPropDL,wolfle2002enhanced,locRtEnt,MLWinPropMultPath}.

It should also be mentioned that the industry state of the art on radio localization as implemented by systems such as Google Maps and Apple are based on a multitude of sensor data, including GNSS, ``radio signatures’’ of all sort (e.g., WiFi SSIDs, cellular channel quality, detailed channel frequency response of the fading channels over the OFDM subcarriers, and so on), and these data are fused via some machine learning scheme. Since these practical implementations are ad-hoc industry intellectual property and not revealed in the open literature at a level of details sufficient to run accurate performance comparisons, it is virtually impossible to assess the performance of our proposed method against the ``real-world’’ state of the art. In contrast, we hasten to say here that our method is based uniquely on received signal level measurements from known BSs. These measurements are collected by default by standard user devices since each device continuously measures the received signal strength from BS beacon signals, in order to select the best cell and trigger handovers. Furthermore, these measurements are very robust since BS beacon signals are easy to detect and do not require any calibration or accurate timing and frequency synchronization, since they are simply wideband power measurements. 

Instead of traditional simulations, in our approach we use RadioUNet \cite{levie2019radiounet} to approximate pathloss functions. RadioUNet is a deep learning simulator, which computes the pathloss function from the map of the city, location of BSs, and optionally some measurements of the true pathloss at some known locations. When only the map and BSs are given as inputs, the method is called RadioUNet$_C$, and when some ground truth measurements are given the method is called RadioUNet$_S$. RadioUNet is orders of magnitude faster than physical simulations, with or without input measurements. Moreover, RadioUNet$_S$ is less sensitive to inaccuracies in the domain, since it is a hybrid between a physical simulation and interpolation of the pathloss measurements.

To the best of our knowledge, the most similar idea to ours was proposed in \cite{limitsRSS} and again recently in \cite{mapBased}. In \cite{mapBased}, the authors consider three different antennas and generate their signal strength maps by extensive measurements. Based on the reported signal strength of the device to be located, the authors propose to take the intersection of the contour maps. As opposed to that method, our method does not rely on expensive measurement campaigns, works in any environment, and does not require any tuning to fit each specific environment. Moreover, instead of intersecting level-sets we consider their sum, which significantly improves the stability of the method.

When using RadioUNet$_S$, the measurements either come from other BS, for which we know the locations, or from other UE with known locations. We justify the latter case as follows: There are many localization techniques based on different fingerprints. Different methods fail at different situations, and it is thus beneficial to rely on a set of methods for localization. Hence, we suppose that a subset of UEs can be localized using some other methods. These devices, which have known locations, report the received pathloss from each BS, and are hence used as input measurements in RadioUNet$_S$ to localize the devices for which the other methods failed.

\textbf{Our contribution:} Our contribution can be summarized as follows:
\begin{itemize}
\item We propose a computationally efficient localization method, merely based on RSS measurements, which does not necessitate expensive hardware at the devices.

    \item The presented method relies on radio map (pathloss function) estimations and the reported received signal strength (RSS) values from the device of interest. Given the transmit powers of the devices, we can compute the pathloss experienced by the UE. Given the found pathloss of UE, we estimate the location of UE by comparing with the radio map. By considering RSS reports at multiple BSs, we propose a simple voting idea to pinpoint the location of the UE.
    
     \item Using the recently developed RadioUNet, we can estimate radio maps very efficiently and accurately, which was not possible until very recently. The RadioUNet deals with uncertainties in the environment via reported measurements from nodes with known locations, e.g., the BSs or other devices with known locations.  

\end{itemize}

\section{RadioUNet}
\label{RadioUNet}

RadioUNet is a deep learning-based simulation method introduced in  \cite{levie2019radiounet}. The first version of the method, RadioUNet$_C$, is a function that receives the map of the city and the location of a BS and returns an estimation of the corresponding radio map. The second version of the method, RadioUNet$_S$, is a function that receives the map of the city, the location of a BS, and a list of measurements of the ground truth radio map at some locations, and returns an estimation of the radio map. Both method achieve high accuracy, with root-mean-square-error of order of 1\% in various scenarios, and a run-time order of milliseconds on NVIDIA Quadro GP100.

RadioUNet is trained in supervised learning to match simulations of radio maps, using the RadioMaSeer dataset\footnote{\url{https://RadioMapSeer.github.io/}}. The dataset consists of 56,000 coarse simulations, that approximate real-life radio maps roughly. RadioMapSeer also contains a smaller dataset of 1400 simulated radio maps using high accuracy simulations. The high accuracy simulations serve as surrogates to real-live radio maps for testing. Moreover, the high accuracy simulations are used for training as explained next.

Since the coarse simulations only roughly approximate real-life radio maps, it is important to adapt and modify what was learned from the coarse simulations to real life. It is assumed that a small dataset of real-life measurements are given, where in each of the 1400 simulations only a small number of locations are sampled, instead the full radio map on the dense grid. Namely, we consider a realistic measurement campaign. In the RadioMapSeer dataset, measurements from the high accuracy simulations are given instead of real-life measurements.

RadioUNet is first trained to match the coarse simulations. After this first training, RadioUNet is optionally further trained, by tuning a relatively small number of trainable parameters, to match the measured high accuracy simulated dataset. In practice, this second step is performed with real-life measurements, adapting what was learned from the coarse simulations to real-life.

In all of the above settings, a second version of each simulation is given, where the effects of cars, randomly generated near and along the streets, is also simulated.

\section{Pathloss Based fingerprint Localization}

In this section we present our pathloss based approach for localization.

\subsection{Level-set intersection localization}\label{subseq:LvLInters}

Suppose a device reports the strength of pilot signals, transmitted from some BSs with known locations. Based on the relation between transmit/receive powers and the pathloss
 $(\textup{P}_{\textup{Rx}})_{\rm dB} = \textup{P}_{\textup{L}}+ (\textup{P}_{\textup{Tx}})_{\rm dB}$
 the pathloss between the device and the BSs can be calculated, where we assume the small-scale fading effects are eliminated by time averaging \cite{vehicle69}. In our approach, a central processor (CP), to which the BSs are connected, estimates the position of the device based on the pathloss values and estimations of the radio maps, which the CP also computes.

 In this situation, if we know the radio maps of the BSs we can in general localize the device accurately. Indeed, given that the signal from BS $S_k$ experienced a pathloss $g_k$, the location $x$ or the device can only be in the set
\[ L_k= \{z\ |\ r_k(z)= g_k\},\]
where $r_k:\mathbf{G}\rightarrow\mathbb{R}$ is the radio map of station $S_k$, and $\mathbf{G}$ is a 2D  grid of measurements. We call $L_k$ a level-set of $r_k$.

Assuming some regularity of $r_k$, the set $L_k$ is one dimensional in the generic case (a union of curves). Thus, in the general case, it is sufficient to consider two BSs $S_1$ and $S_2$, since we require
\[x\in L_1\cap L_2,\]
and $L_1\cap L_2$ is zero dimensional in general (a sparse collection of points). Using three BSs, $L_1\cap L_2\cap L_3$ gives almost always one point. 

However, we only have access to approximations $\tilde{r}_k$ of $r_k$. We thus relax the intersection method, replacing $L_k$ by
\[ L^{\epsilon}_k= \{z\ |\ |\tilde{r}_k- g_k|<\epsilon\},\]
for some small $\epsilon$. We call  $L^{\epsilon}_k$ an \textit{$\epsilon$-level-set} of $\tilde{r}_k$.

We also consider $N\geq 3$ base-stations. 
One simple way to obtain a small set in which the UE is located is to intersect the $\epsilon$-level-sets.
For this, define the indicator function of the  $\epsilon$-level-sets 
 \[
    l^{\epsilon_k}_k(z) = \left\{\begin{array}{lr}
        1, & \text{if } z\in L^{\epsilon_k}_k\\
        0, & \text{if } z\notin L^{\epsilon_k}_k\\
        \end{array}\right.
  \]
The intersection of the $\epsilon$-level-sets is now given by the support of the product function
\begin{equation}
\prod_{k=1}^K l^{\epsilon_k}_k.
\label{prod1}
\end{equation}
Note that each BS can have a different $\epsilon_k$.

To the best of our knowledge, this method was first presented in \cite{limitsRSS} under the name \textit{Simple Point Matching} (SPM). Notice that intersections may result in empty sets. To obtain non-empty intersections, the authors start with low $\epsilon_k = \epsilon$ for all $k$ and gradually increase it until a non-zero intersection set is found. 

\subsection{Level-set voting method}

Note that even though the RMSE error between $\tilde{r}_k$ and $r_k$ is small, the pointwise error can still be large at a small number of locations. If $x$ happens to be a location where $|\tilde{r}_k(x)-r_k(x)|>\epsilon$, then $x$ is not located in the $\epsilon$-level-set of $\tilde{r}_k$, and the intersection of all $\epsilon$-level-sets will generally be an empty set and fail to localize the UE. 

To solve this problem, we replace the level-set intersection method with a \emph{level-set voting method}. The basic idea is to replace the product in (\ref{prod1}) by the sum
\begin{equation}
V(z)=\sum_{k=1}^K l_k(z).
\label{vote1}
\end{equation}
The value of $V(z)$ of (\ref{vote1}) at each location $z$ is interpreted as the number of votes supporting $z$ as the location of the UE. Here, each BS votes 1 for the locations in which the UE can be located according to the reported signal strength from the BS, and votes zero elsewhere. Namely, each BS $k$ votes one in $L^{\epsilon_k}_k$ and zero outside $L^{\epsilon_k}_k$. Summing all votes for each location is equivalent to taking the sum in (\ref{vote1}). Hence, the set of points $z$ with maximal $V(z)$ value are taken as the localization set of UE. We denote this set by $O$.

To obtain a 2D location for UE, instead of a set, the center of mass of $O$ is computed
by
\[X= \sum_{z\in O}  \frac{z }{|O|},\]
where $|O|$ is the area of $O$ given by
$\sum_{z\in O}  1$.

High quality localization outcomes are those for which $O$ is localized about $X$.
Thus, we asses the quality/reliability of the localization outcome $X$ by the standard deviation
\[Q= \sqrt{\sum_{z\in O}  \frac{|z-X|^2}{|O|} }.\]

\section{Numerical results}
\subsection{Examples}
In the following examples we consider various versions of ground truth radio maps. The UE reports pathloss sampled at its location based on the assumed ground truth. In the voting method, the estimations of the radio maps are computed using various RadioUNet methods.
We consider map 316 from the test set of RadioMapSeer.

\begin{figure}[!htb]
	\centering
	\begin{subfigure}{0.24\textwidth}\centering
		\includegraphics[width=0.9\linewidth]{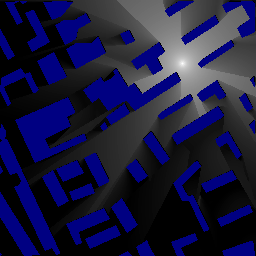}
		\caption{\footnotesize DPM example ground truth \newline}
		\label{DPMTruth}
	\end{subfigure}%
	\hfill
	\begin{subfigure}{0.24\textwidth}\centering
		\includegraphics[width=0.9\linewidth]{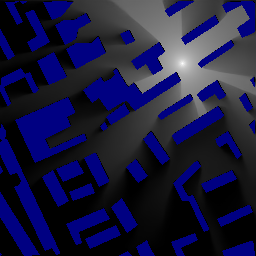}
		\caption{\footnotesize DPM example estimation  with RadioUNet$_C$}
		\label{DPMEst}
	\end{subfigure}%
	\hfill
	\begin{subfigure}{0.24\textwidth}\centering
		\includegraphics[width=0.9\linewidth]{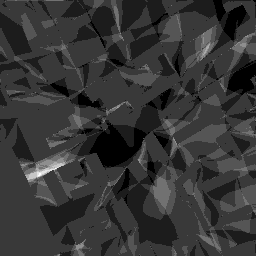}
		\caption{\footnotesize DPM level sets sum  \newline \newline}
		\label{DPMLvl}
	\end{subfigure}%
	\hfill
	\begin{subfigure}{0.24\textwidth}\centering
		\includegraphics[width=0.9\linewidth]{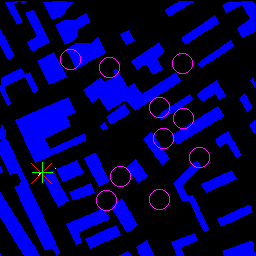}
		\caption{\footnotesize DPM localization \newline \newline}
		\label{DPMLoc}
	\end{subfigure}%
	\caption{ DPM localization results. Distance between the estimated and ground truth UE location: 1.3m. $Q=4.2$m. \newline}
	\label{fig:DPM}
\end{figure}

\begin{figure}
	\centering
	\begin{subfigure}{0.24\textwidth}\centering
		\includegraphics[width=0.9\linewidth]{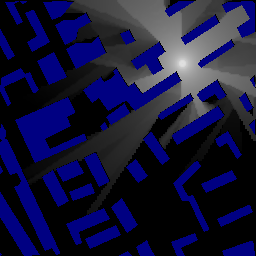}
		\caption{\footnotesize IRT2 example ground truth \newline}
		\label{IRT2Truth}
	\end{subfigure}%
	\hfill
	\begin{subfigure}{0.24\textwidth}\centering
		\includegraphics[width=0.9\linewidth]{I40predictB.png}
		\caption{\footnotesize IRT2 example estimation as DPM with RadioUNet$_C$}
		\label{IRT2Est}
	\end{subfigure}%
	\hfill
	\begin{subfigure}{0.24\textwidth}\centering
		\includegraphics[width=0.9\linewidth]{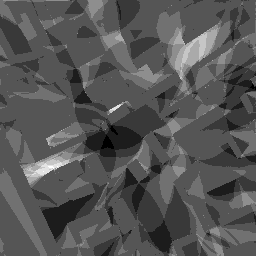}
		\caption{\footnotesize IRT2 level sets sum  \newline \newline}
		\label{IRT2Lvl}
	\end{subfigure}%
	\hfill
	\begin{subfigure}{0.24\textwidth}\centering
		\includegraphics[width=0.9\linewidth]{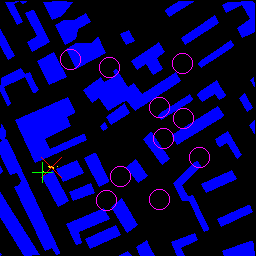}
		\caption{\footnotesize IRT2 localization \newline \newline}
		\label{IRT2Loc}
	\end{subfigure}%
	\caption{IRT2 localization results. Distance between the estimated and ground truth UE location: 10m. $Q=2.4$m. \newline}
	\label{fig:IRT2}
\end{figure}

\begin{figure}
	\centering
	\begin{subfigure}{0.24\textwidth}\centering
		\includegraphics[width=0.9\linewidth]{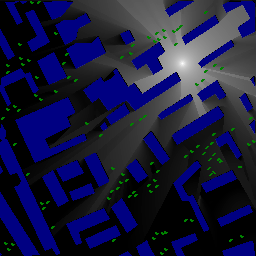}
		\caption{\footnotesize DPM with cars example ground truth \newline}
		\label{DPMcarsTruth}
	\end{subfigure}%
	\hfill
	\begin{subfigure}{0.24\textwidth}\centering
		\includegraphics[width=0.9\linewidth]{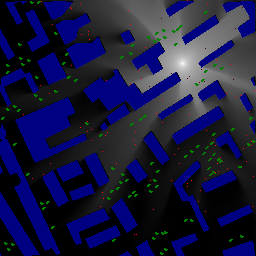}
		\caption{\footnotesize DPM with cars example estimation with  RadioUNet$_S$}
		\label{DPMcarsEst}
	\end{subfigure}%
	\hfill
	\begin{subfigure}{0.24\textwidth}\centering
		\includegraphics[width=0.9\linewidth]{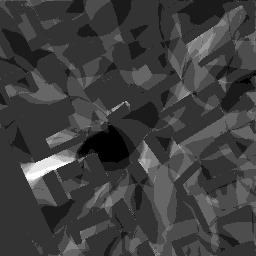}
		\caption{\footnotesize DPM with cars level sets sum \newline}
		\label{DPMcarsLvl}
	\end{subfigure}%
	\hfill
	\begin{subfigure}{0.24\textwidth}\centering
		\includegraphics[width=0.9\linewidth]{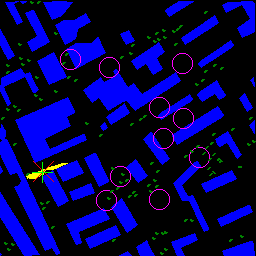}
		\caption{\footnotesize DPM with cars localization \newline}
		\label{DPMcarsLoc}
	\end{subfigure}%
	\caption{DPM with cars. Input measurements of RadioUNet$_S$ marked in red. Cars marked in green. Distance between the estimated and ground truth UE location: 1.6m. $Q=8.5$m. \newline}
	\label{fig:DPMcars}
\end{figure}

\begin{figure}
	\centering
	\begin{subfigure}{0.24\textwidth}\centering
		\includegraphics[width=0.9\linewidth]{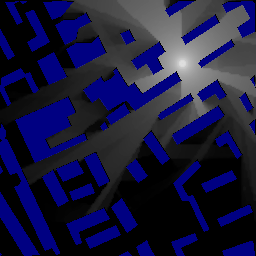}
		\caption{\footnotesize IRT4 example ground truth \newline}
		\label{IRT4Truth}
	\end{subfigure}%
	\hfill
	\begin{subfigure}{0.24\textwidth}\centering
		\includegraphics[width=0.9\linewidth]{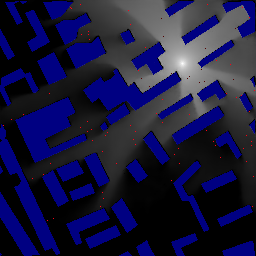}
		\caption{\footnotesize IRT4 example estimation with RadioUNet$_S$}
		\label{IRT4Est}
	\end{subfigure}%
	\hfill
	\begin{subfigure}{0.24\textwidth}\centering
		\includegraphics[width=0.9\linewidth]{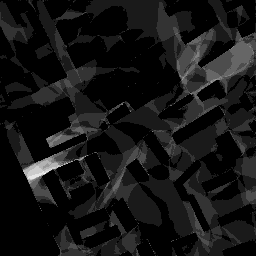}
		\caption{\footnotesize IRT4 level sets sum  \newline \newline}
		\label{IRT4Lvl}
	\end{subfigure}%
	\hfill
	\begin{subfigure}{0.24\textwidth}\centering
		\includegraphics[width=0.9\linewidth]{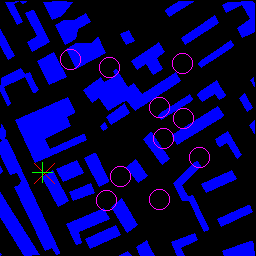}
		\caption{\footnotesize IRT4 localization \newline \newline}
		\label{IRT4Loc}
	\end{subfigure}%
	\caption{High accuracy simulation (IRT4) localization results. Input measurements of RadioUNet$_S$ marked in red. Distance between the estimated and ground truth UE location: 2.24m. $Q=0$m. \newline}
	\label{fig:IRT4}
\end{figure}

In Figures \ref{fig:DPM},\ref{fig:IRT2},\ref{fig:DPMcars},\ref{fig:IRT4}, BSs are marked with magenta circles. UE true and estimated location are marked with green cross and red X, respectively. Localization set $O$ is denoted by yellow.

We first consider RadioUNet$_C$ trained on coarse simulations and tested against coarse simulations. The coarse simulation we use is called DPM (dominant path model). The results are presented in Figure \ref{fig:DPM}. 

Next, we use RadioUNet$_C$, trained only against coarse simulations (DPM), but tested against a different type of simulation called IRT2 (intelligent ray tracing with 2 ray interactions). The idea is that the DPM coarse simulations, used for training, are quite different from IRT2. This setting mimics the scenario where our estimation does not capture real-life radio maps well, and the learned method was not adapted to a dataset of real-life measurements. Hence, we see these results as worse case estimations, bounding the general performance of our method. The results, presented in Figure \ref{fig:IRT2}, showcase that our voting method is relatively robust and  performs reasonably also on low quality radio map estimations. 

We then consider RadioUNet$_S$, trained on a dataset of simulations including the effects of cars with unknown locations. The method compensates for the uncertainty in the domain (the unknown car locations) by using the small number of ground truth measurements (200 measurements). The results are reported in Figure \ref{fig:DPMcars}.

Last, we consider RadioUNet$_S$, trained on coarse simulations and adapted to measured high accuracy simulation, as explained in Section \ref{RadioUNet}. The ground truth reported pathlosses are taken from the high accuracy simulations, but the radio maps used in the voting method are computed using  RadioUNet$_S$. The results are reported in Figure \ref{fig:IRT4}.

\FloatBarrier

\vspace{-4mm}

\vspace{-4mm}	

\subsection{Statistics and Tuning}

\begin{figure}[!ht]
\vspace{-2mm}
    \centering
     \includegraphics[width=0.75\linewidth]{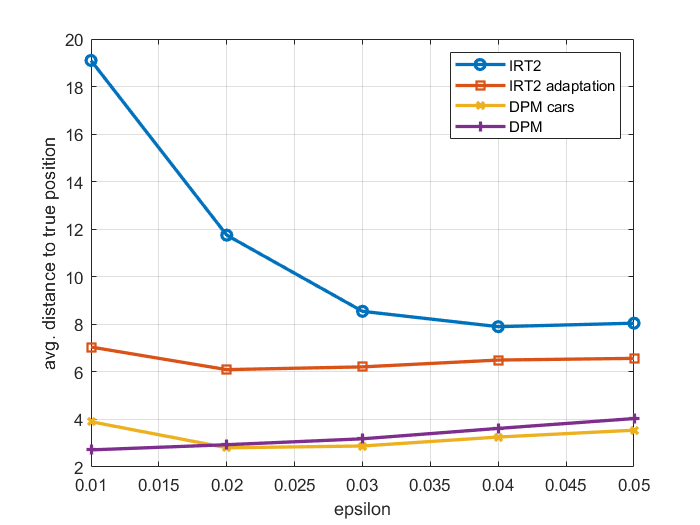}

    \vspace{-2mm}
    \caption{Average estimation error versus $\epsilon$.}
    \label{fig:diffVsEps}
\end{figure}


\begin{figure}[!ht]
\vspace{-2mm}
    \centering
     \includegraphics[width=0.75\linewidth]{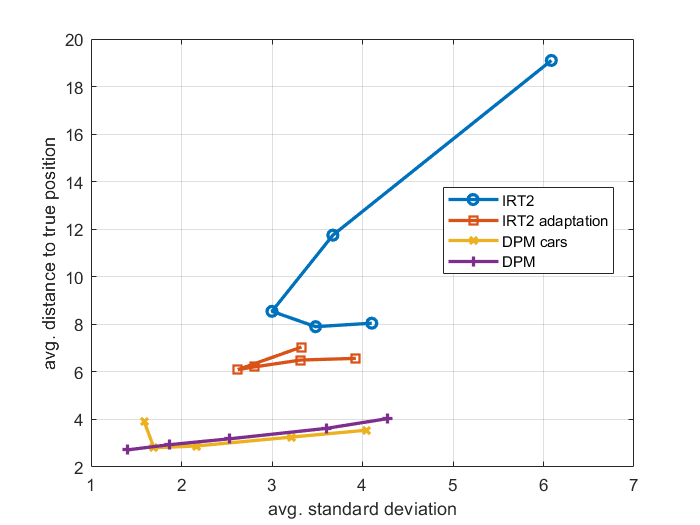}

    \vspace{-2mm}
    \caption{Average standard deviation vs average accuracy for $\epsilon \in \{0.01,0.02,0.03,0.04,0.05 \}$.}
    \label{fig:diffVsStd}
\end{figure}

\begin{figure}[!ht]
\vspace{-2mm}
    \centering
     \includegraphics[width=0.75\linewidth]{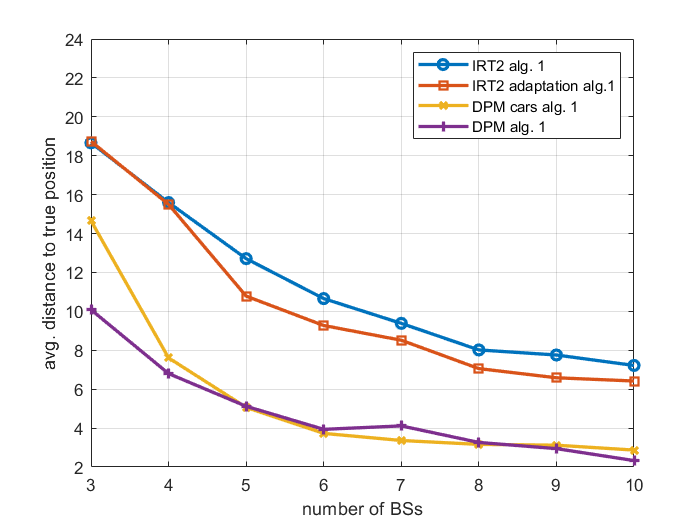}

    \vspace{-2mm}
    \caption{Average estimation error versus number of BSs for Algorithm 1.}
    \label{fig:alg1}
\end{figure}

\begin{figure}[!ht]
\vspace{-2mm}
    \centering
     \includegraphics[width=0.75\linewidth]{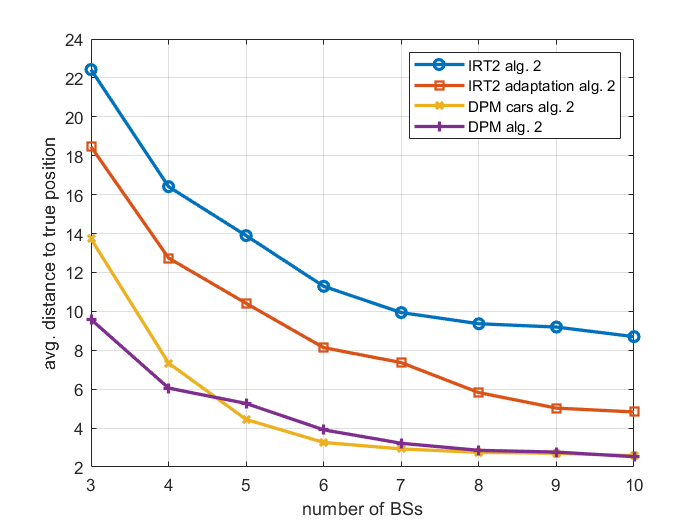}

    \vspace{-2mm}
    \caption{Average estimation error versus number of BSs for Algorithm 2.}
    \label{fig:alg2}
\end{figure}

We consider a dataset of radio maps, the 80 maps from the test set of RadioMapSeer. These maps are randomly split to 40 training and 40 test maps. We compute the following statistics on the training maps. For each value of $\epsilon$ in a grid between 0.01 and 0.05 with step size 0.01, we compute the localization outcomes in all training maps, and plot the average localization accuracy as function of epsilon (see Fig. \ref{fig:diffVsEps}, where 10 BSs were considered). 

We consider two algorithms based on the resulting statistics. In the first algorithm, we choose the $\epsilon$ that minimizes the localization error in the training set, and use this fixed $\epsilon$ in the voting algorithm.
For the second algorithm, in a precomputation step, we plot the average localization accuracy as a function of the average localization standard deviation $Q$ (averaged over the training set). We pick the optimal standard deviation that minimizes the localization error (see Fig. \ref{fig:diffVsStd}, where 10 BSs were considered). Then, in the localization algorithm we produce 5 localization-outcomes based on 5 different $\epsilon$ values between 0.01 and 0.05. Out of these, we pick the outcome with standard deviation closest to the optimal standard deviation. 

We evaluate both algorithms on the test set.  
We consider the following scenarios:  First, RadioUNet$_C$ trained on coarse DPM simulations and tested against coarse DPM simulations. Second, RadioUNet$_C$ trained on coarse DPM simulations and tested on IRT2 simulations. Third, RadioUNet$_S$, trained on a dataset of simulations including the effects of cars with unknown locations. We also consider RadioUNet$_C$, trained on coarse DPM  simulations and adapted to a small dataset of measured IRT2 simulations, as explained in Section \ref{RadioUNet}. 
The resulting accuracies for the first and second algorithms are shown in Figures \ref{fig:alg1} and \ref{fig:alg2}, respectively, as a function of number of BSs.
Note that the DPM method which is not adapted to IRT2 represents a worst case bound on the accuracy of our method, since we did nothing to adapt the method from simulation (DPM) to real-life deployment (represented by IRT2). The DPM method which is adapted to measured IRT2 performs much better.

\section{Conclusions}

In this paper we presented a simple but effective way to localize a UE given reported signal strengths from a set of BSs. Our method, based on estimating the radio maps of each BS, works in realistic propagation environments in urban situation. The proposed approach does not necessitate expensive hardware, since the estimations of the radio maps are computed very efficiently via RadioUNets. Moreover, it can be easily integrated in the localization methods that are already in use.

\section{Acknowledgment}

The work presented in this paper was partially funded by the DFG Grant DFG SPP 1798 “Compressed Sensing in Information Processing” through Project Massive MIMO-II, and by the German Ministry for Education and Research as BIFOLD - Berlin Institute for the Foundations of Learning and Data (ref. 01IS18037A).




\bibliographystyle{IEEEtran}
\bibliography{pub}

\end{document}